# Light Transmission Through Tow Dimensional Subwavelenght Magneto-Optical Slit Array

PAYMAN PAHLAVAN

**Abstract:** In this paper, metallic behavior of magneto optic materials in slit arrays under the incident of TM wave is analyzed by mode matching technique and confirmed by full wave simulations. It is shown that tensor electric permittivity of such materials under a static magnetic field bias can result in negative effective permittivity and if placed in a periodic two dimensional structure known as slit array, results in extraordinary optical transmission. It is also shown that this structure acts as a highly frequency selective surface tunable with magnetic bias change.

## 1. Introduction

Despite the previously believed idea of low transmission efficiency of light through subwavelength apertures [1], Ebbesen et al showed that a 2D periodic array in a metallic film exhibits an extraordinary optical transmission (EOT) even at the wavelengths ten times bigger than the hole dimension [2]. The excitation of surface plasmon polaritons (SPPs) was then suggested to have a crucial role in this phenomenon [3]. Following this idea, intensive studies have been later done to expand this topic mostly due to its potential applications in important areas such as light localization, microcavity quantum electrodynamics, near-field optics photolithography, and light extraction from LEDs [4-8]. Later on, as a simpler alternative structure, researchers have further investigated one-dimensional narrow slit array having similar light transmission behavior with similar practical applications mentioned above [9]. It has been shown that the excitation of the surface electromagnetic waves, caused by the incident light, yields in an extraordinary light transmission in the forward direction [10]. Afterwards, analytical studies implied that the diffraction and composition of evanescent waves occur at EOT frequencies too [11]. The infinite permittivity of a perfect metal which plays the main role in this optical transmission can also be mimicked by magneto optic materials in a specific frequency region.

In general, magneto optic materials, under a static magnetic field bias, represent a tensor permittivity in terahertz and optical frequencies, similar to how ferrites, as the most known magnetic materials; behave in their permeability constant in radio frequencies.

Based on Drude-Zener model of semiconductors [12], a semiconductor magnetized in z-direction possesses a relative tensor permittivity. This behavior is due to cyclotron motion of the carriers around the direction of the applied magnetic field. In these materials, the relative permittivity can be generally expressed as

$$\bar{\bar{\varepsilon}} = \varepsilon_0 \begin{pmatrix} \varepsilon_1 & -j\varepsilon_2 & 0 \\ -j\varepsilon_2 & \varepsilon_1 & 0 \\ 0 & 0 & \varepsilon_3 \end{pmatrix} \qquad (1)$$

$$\varepsilon_1 = \varepsilon_\infty - \frac{\omega_p^2(\omega^2 + j\gamma\omega)}{(\omega^2 + j\gamma\omega)^2 - \omega^2\omega_c^2}, \quad \varepsilon_2 = \frac{\omega\omega_c\omega_p^2}{(\omega^2 + j\gamma\omega)^2 - \omega^2\omega_c^2}, \quad \varepsilon_3 = \varepsilon_\infty - \frac{\omega_p^2}{(\omega^2 + j\gamma\omega)^2} \qquad (2)$$

here $\omega_p \sim 10^{13}$ s$^{-1}$ and $\omega_c \simeq 0.5\omega_p$ are the plasma and cyclotron angular frequency respectively, $\frac{\gamma}{2\pi} = 0.05$THz is the collision frequency, N is density of "nearly free" electrons in semiconductor, m* is the effective electron mass, $B_0 \sim \frac{\omega_c}{\omega_p}$ is the applied magnetic flux, e is the electron charge, and $\epsilon_\infty \simeq 15.68$ is the high frequency background dielectric constant For Indium-Antimonide (InSb)[12]. The effective permittivity $\epsilon_\perp = \epsilon_1 - {\epsilon_2}^2/{\epsilon_1}$ is defined in electromagnetic field propagation in the particular case transverse magnetic (TM) waves inside a magneto optic material [13]. It is shown in Fig. 1 that the real part of this value is highly negative in frequencies bellow 300GHz, making this material a suitable substitution for metals.

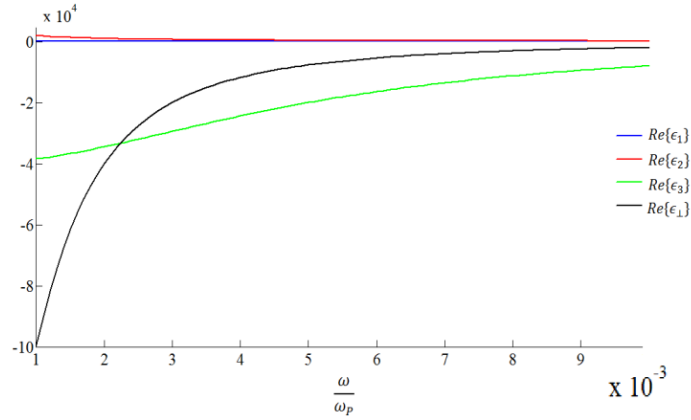

Fig. 1: Real parts of permittivitivy tensor elemets of InSb versus normalized frequency.

In this manuscript, we analytically study very narrow slit array on InSb using mode matching method and numerically investigate them employing full-wave simulation techniques such as FEM and FIT. In section 2, we derive EM modes and the dispersion equations of slit array on an In-Sb film, and in section 3, we calculate the EM transmission through a two dimensional array of Air-InSb slits and show the dependence of the transmission frequencies on the external static magnetic field, making this structure an ultra-narrow tunable frequency selective filter.

## 2. Slit Array Eigen modes

An infinite air slit array created on an In-Sb film can be treated as an array of two dimensional waveguides. In this periodic structure shown in Fig. 2, with no variation along z-axes ($\frac{\partial}{\partial z} = 0$), Maxwell equations for TM waves inside the InSb can be written as:

$$\begin{pmatrix} E_x \\ E_y \end{pmatrix} = \frac{1}{j\omega\varepsilon_0(\varepsilon_1^2 - \varepsilon_2^2)} \begin{pmatrix} \varepsilon_1 & j\varepsilon_2 \\ -j\varepsilon_2 & \varepsilon_1 \end{pmatrix} \begin{pmatrix} \frac{\partial H_z}{\partial y} \\ \frac{\partial H_z}{\partial x} \end{pmatrix} \qquad (3)$$

$$\frac{\partial E_x}{\partial y} - \frac{\partial E_y}{\partial x} = -j\omega\mu H_z \qquad (4)$$

Therefore, the Helmholtz equation is

$$\nabla^2_{x,y} H_z + k_0^2 \varepsilon_\perp H_z = 0$$

The effective permittivity $\varepsilon_\perp$ is highly negative in our frequency range, while other elements of the permittivity tensor are positive. It indicates that TM waves in In-Sb cannot propagate. This also indicates that in Air-InSb intersection, surface waves called Magneto-palsmonic waves can propagate. However unlike Palsmonic surface waves, propagation constant value in Magneto-palsmonic surface waves depends on the propagation direction [14]. Another interesting fact is that in negative $\varepsilon_\perp$ frequencies, all the incident wave upon the MO material is reflected, similar to PEC surface wave reflection [13]. Base on Figure 1, in frequencies up to 300GHz, we can find a highly negative region for $\varepsilon_\perp$, and obtain all Eigen modes.

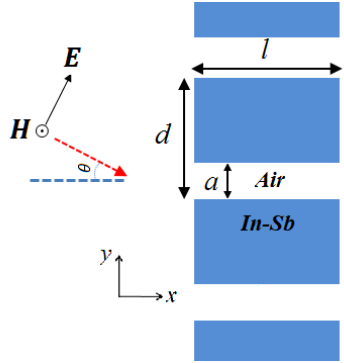

Fig. 2: tow dimentional arry of air slits cut in a InSb film.

Thanks to the permittivity tensor of the biased InSb, the structure illustrated in Fig. 3, is asymmetric. Therefore, unlike other structures, electromagnetic waves cannot be decoupled into even and odd modes, but a combination of them should be considered in field distribution. Hence, transverse Electric and Magnetic field in both media can be defined as

$$H_z = e^{-j\beta x} \begin{cases} A\sinh\alpha_0 y + B\cosh\alpha_0 y & 0 < y < a \\ Ce^{-\alpha y} + De^{\alpha y} & a < y < d \end{cases} \quad (5)$$

$$E_x = \frac{e^{-j\beta x}}{j\omega\varepsilon_0} \begin{cases} A\alpha_0\cosh\alpha_0 y + B\alpha_0\sinh\alpha_0 y & 0 < y < a \\ C\dfrac{(-\varepsilon_1\alpha - \varepsilon_2\beta)}{\varepsilon_1\varepsilon_\perp}e^{-\alpha y} + D\dfrac{(\varepsilon_1\alpha - \varepsilon_2\beta)}{\varepsilon_1\varepsilon_\perp}e^{\alpha y} & a < y < d \end{cases} \quad (6)$$

$$E_y = \frac{e^{-j\beta x}}{\omega\varepsilon_0} \begin{cases} A\beta\cosh\alpha_0 y + B\beta\sinh\alpha_0 y & 0 < y < a \\ C\dfrac{(\varepsilon_2\alpha + \varepsilon_1\beta)}{\varepsilon_1\varepsilon_\perp}e^{-\alpha y} + D\dfrac{(-\varepsilon_2\alpha + \varepsilon_1\beta)}{\varepsilon_1\varepsilon_\perp}e^{\alpha y} & a < y < d \end{cases} \quad (7)$$

$$\alpha_0 = \sqrt{\beta^2 - k_0^2}, \quad \alpha = \sqrt{\beta^2 - k_0^2 \varepsilon_\perp} \quad (8)$$

This asymmetry is obvious in Electric field patterns based on equations (7) and (8). Removing the external bias turns the InSb to a homogeneous medium by vanishing $\varepsilon_2$. Under such condition, changing the sign of beta does not change the electric field patterns.

Applying boundary and periodicity conditions in $y = a$ and $y = 0, d$ and defining A and B as a function of C and D result in a matrix equation as

$$\begin{pmatrix} e^{-\alpha(d-a)}\left(\cosh\alpha_0\alpha - f\sinh\alpha_0\alpha\right) - 1 & e^{\alpha(d-a)}\left(\cosh\alpha_0\alpha + g\sinh\alpha_0\alpha\right) - 1 \\ e^{-\alpha(d-a)}\left(\sinh\alpha_0\alpha - f\cosh\alpha_0\alpha\right) + f & e^{\alpha(d-a)}\left(\cosh\alpha_0\alpha + g\sinh\alpha_0\alpha\right) - g \end{pmatrix}\begin{pmatrix} C \\ D \end{pmatrix} = 0 \quad (9)$$

$$f = \frac{\varepsilon_1\alpha + \varepsilon_2\beta}{\alpha_0\varepsilon_1\varepsilon_\perp}, \quad g = \frac{\varepsilon_1\alpha - \varepsilon_2\beta}{\alpha_0\varepsilon_1\varepsilon_\perp} \quad (10)$$

For this equation to be always zero regardless of C and D, the matrix determinant should always be zero. Therefore, the dispersion relation of Electromagnetic waves in this structure is expressed as

$$\left(\frac{2\varepsilon_1\alpha}{\alpha_0\varepsilon_1\varepsilon_\perp}\right)\left(2 - \left(e^{-\alpha(d-a)} + e^{\alpha(d-a)}\right)\cosh(\alpha_0\alpha)\right) + \left(\frac{\varepsilon_1^2\alpha^2 - \varepsilon_2^2\beta^2}{(\alpha_0\varepsilon_1\varepsilon_\perp)^2} + 1\right)\left(e^{-\alpha(d-a)} - e^{\alpha(d-a)}\right)\sinh(\alpha_0\alpha) = 0 \quad (11)$$

Equation (9) has trivial solution if the determinant is zero. Hence A, B and C can be easily calculated by choosing D=1. Transverse magnetic field distribution is illustrated in Fig. 4. As it was anticipated, field amplitude is decaying in InSb due to high negative permittivity.

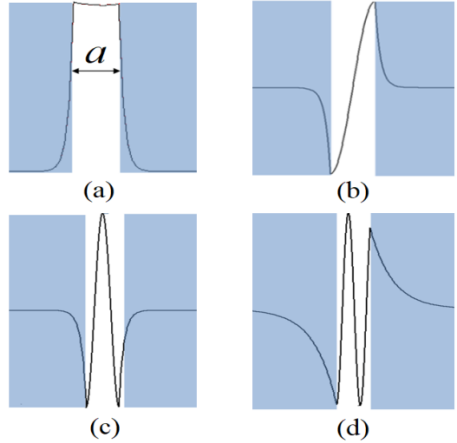

Fig. 4: Transverse magnetic field distribution for first four modes of Air-InSb array.

Similar to two dimensional slit arrays on PEC, In order to investigate the EOT in this structure, the period constant $d$ must always be smaller than the wavelength of the incident wave. In this case, higher order diffracted waves, Floquet-Bloch higher harmonic modes, are highly evanescent and the transmission only occurs for the first harmonic of the incident wave.

Base on fig 2. By choosing $\lambda > d = 1mm$, the effective permittivity will be smaller than -500 which is a high negative number. Therefore, InSb behaves like a metal but this behavior changes with external static magnetic field bias.

The air slit width $a = 0.1d$ is also defined so that just one mode is propagating along the slit while higher modes are highly evanescent. With just one propagating mode inside the structure and one propagating diffracted mode, mode matching technique with single mode approximation results in good agreement with simulation

### 3. Transmission

One-dimensional array of narrow slits on the In-Sb film, which is surrounded by free space, is shown in Fig. 3. This is a periodic structure in *x*-direction with period *d*, slit width *a*, and *l* as the thickness. The EOT only appears when a TM polarized wave is incident upon this structure even with very narrow slits [12]. For the sake of simplicity, only the normal incident is studied here. Here free spaces files are denoted by index 0 and slit modes are indicated by index 1. Considering Floquet-Bloch modes, all the components of the incident, reflected, and transmitted waves in both interfaces can be written as [13]:

$$H_z^0\big|_{x<0} = \exp[-jk_0 x] + \sum_n \rho_n \exp[jk_{x,n}x - jk_{y,n}y]$$
$$H_z^0\big|_{x>l} = \sum_n t_n \exp[-jk_{x,n}x - jk_{y,n}y],$$
(12)

$$E_y^0\big|_{x<0} = \exp[-jk_0 x] - \sum_n \rho_n k_{x,n} \exp[jk_{x,n}x - jk_{y,n}y]$$
$$E_y^0\big|_{x>l} = \frac{1}{\omega\varepsilon_0}\sum_n t_n \exp[-jk_{x,n}x - jk_{y,n}y](\hat{x}k_{y,n} + \hat{y}k_{x,n})$$
(13)

Also the wave number components are defined as:

$$k_{y,n} = 2n\pi/d, \quad k_{x,n} = \sqrt{k_0^2 - k_{y,n}^2}$$
(14)

The tangential slit filed patterns at two interfaces can be written as a sum of forward and backward waves for all slit modes indicated with *m* index.

$$E_y^1\big|_{x=0} = \sum_{m=0}^{+\infty} a_m E_{m,y}^f + \sum_{n=-\infty}^{n=+\infty} b_m E_{m,y}^b$$
$$E_y^1\big|_{x=l} = \sum_{m=0}^{+\infty} a_m E_{m,y}^f e^{-j\beta_m l} + \sum_{n=-\infty}^{n=+\infty} b_m E_{m,y}^b e^{j\beta_m l}$$
and (15)
$$H_z^1\big|_{x=0} = \sum_{m=0}^{+\infty} H_{m,z}(a_m + b_m)$$
$$H_z^1\big|_{x=l} = \sum_{m=0}^{+\infty} H_{m,z}(a_m e^{-j\beta_m l} + b_m e^{j\beta_m l})$$

Due to the periodicity of the structure in y direction, diffracted free space modes possess a wave constant $k_{y,n}$ parallel to the plane of incidence. Hence the perpendicular wave constant $k_{x,n}$ for higher order modes will be imaginary for $\lambda > d/n$. Since the wavelength is always larger than the period, all higher order modes are non-propagating. However their contribution in a precise mode matching in the plane of incidence is significant since it is done by integration of all modes in y direction for one period.

Since the free space zeroth order is the only propagating mode, we can determine the reflection and transmission of this mode by applying the boundary condition in $x = 0$ and $x = d$ interfaces by multiplying both sides of equations with $e^{+jk_0 x}$ and integrating over one period.

$$\int_0^d \left(H_z^0 - H_z^1\right)e^{+jk_0 x}dy = 0$$
$$\int_0^d \left(E_y^0 - E_y^1\right)e^{+jk_0 x}dy = 0 \tag{16}$$

These equations are solved numerically and the results are compared with simulation. In order to solve the equations, only the first slit mode in taken into account while up to 10 free space Floquet-Bloch mode are considered. Fig. 5 shows a good agreement between mode matching technique and the Full wave analysis by CST for $l = 1.1d$ as the thickness of the InSb film.

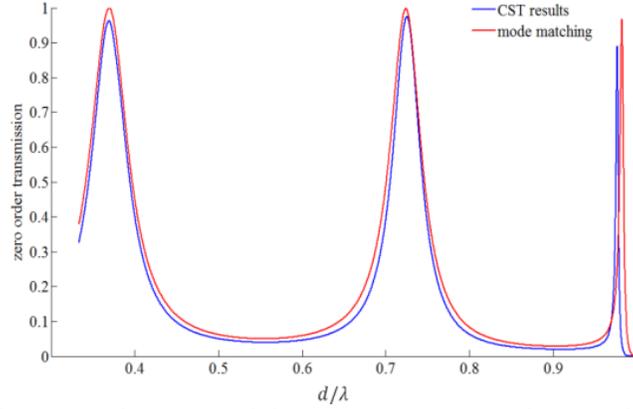

Fig. 5: Electromagnetic wave transmission through subwavelength slit array cut in a InSb film.

The slight difference is due to the material loss defined in CST, while our analysis is done neglecting loss, which appears as an imaginary part in effective permittivity of InSb.

Similar to slit array on PEC film, this structure has two so called Fabry-Perot resonances and an extraordinary transmission at the wavelength near the period value.

It is expected that changing the external magnetic Bias field will change the transmission peak wavelengths because $\omega_c$ and in turn $\varepsilon_\perp$ are a function of Magnetic bias. Different transmissions for several $\omega_c$ values are illustrated in Fig. 6.

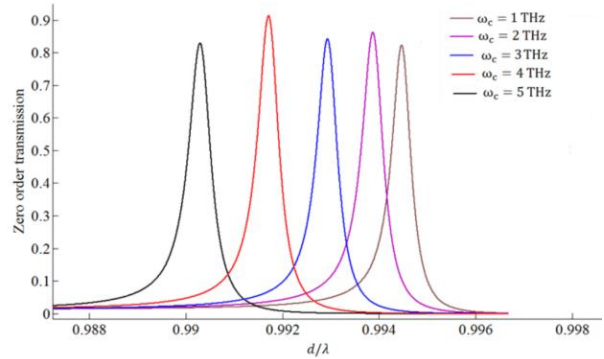

Fig. 6: Electromagnetic wave transmission through subwavelength slit array cut in a narrow InSb film for different $\omega_c$ values.

Base on Fig. 6, a single band frequency selective surface can be designed by decreasing the thickness of the air-InSb periodic structure so that only one transmission peak appears at $\lambda \approx d$ wavelengths. With a thickness $l = 0.1d$, full wave simulations show the anticipated extraordinary transmission which can also be tuned by the external Bias. The effect of the external magnetic bias variation from 0.1 T to 0.5 T on the EOT wavelength is depicted in Fig. 6. In fact, this structure acts as an ultra-narrowband frequency selective surface tunable by an external magnetic bias field.